\documentclass[conference]{IEEEtran}

\usepackage[T1]{fontenc}
\usepackage[latin9]{inputenc}
\usepackage{tikz}
\usetikzlibrary{intersections}
\usetikzlibrary{calc}
\usetikzlibrary{patterns}
\usetikzlibrary{automata}
\usetikzlibrary{positioning}
\usepackage{amsmath}
\usepackage{amssymb}
\usepackage{amsthm}
\usepackage[english]{babel}
\usepackage{cite}

\begin{document}
\title{Reachability in\\Augmented Interval Markov Chains}
\author{\IEEEauthorblockN{Ventsislav Chonev}
\IEEEauthorblockA{IST Austria}}

\newcommand{\lang}{\mathcal{L}} 
\newcommand{\re}{\mathbb{R}} 
\newcommand{\nat}{\mathbb{N}} 
\newcommand{\zed}{\mathbb{Z}} 
\newcommand{\rat}{\mathbb{Q}} 
\newcommand{\thr}{\mathit{Th(\re)}} 
\newcommand{\threxst}{\mathit{Th^{\exists}(\re)}} 
\newcommand{\cclass}[1]{\mathbf{#1}}
\newcommand{\np}{\cclass{NP}}
\newcommand{\pspace}{\cclass{PSPACE}}
\newcommand{\ptime}{\cclass{P}}
\newcommand{\ch}{\cclass{CH}}
\newcommand{\posslp}{\cclass{PosSLP}}
\newcommand{\dblexptime}{\cclass{2}\mbox{-}\cclass{EXPTIME}}
\newcommand{\reach}{\twoheadrightarrow}
\newcommand{\bwreach}{\rightsquigarrow}
\newcommand{\vct}{\mathbf}

\newtheorem{theorem}{Theorem}
\newtheorem{lemma}[theorem]{Lemma}
\newtheorem{proposition}[theorem]{Proposition}
\newtheorem{corollary}[theorem]{Corollary}
\newtheorem{definition}[theorem]{Definition}

\theoremstyle{remark}
\newtheorem{remark}[theorem]{Remark}

\maketitle

\begin{abstract}
In this paper we propose augmented interval Markov chains (AIMCs):
a generalisation of the familiar interval Markov chains (IMCs)
where uncertain transition probabilities are in addition
allowed to depend on one another. This new model preserves the 
flexibility afforded by IMCs for describing stochastic systems 
where the parameters are unclear, for example due to measurement
error, but also allows us to specify transitions with probabilities
known to be identical, thereby lending further expressivity.

The focus of this paper is reachability in AIMCs. We study
the qualitative, exact quantitative and approximate reachability problem, as well as
natural subproblems thereof, and establish several upper and lower 
bounds for their complexity. We prove the exact reachability problem is
at least as hard as the famous square-root sum problem,
but, encouragingly, the approximate version lies in
$\mathbf{NP}$ if the underlying graph is known, 
whilst the restriction of the exact problem to a constant
number of uncertain edges is in $\mathbf{P}$. Finally, we show that
uncertainty in the graph structure affects complexity
by proving $\mathbf{NP}$-completeness for the qualitative subproblem, 
in contrast with an easily-obtained upper bound of $\mathbf{P}$
for the same subproblem with known graph structure.
\end{abstract}

%
\IEEEpeerreviewmaketitle

\section{Introduction}

Discrete-time Markov chains are a well-known stochastic model, one which
has been used extensively to reason about software systems~\cite{mcApplication1,
mcApplication2,mcApplication3}.
They comprise a finite set of states and a set of transitions labelled
with probabilities in such a way that the outgoing transitions from each
state form a distribution. They are useful for modelling systems with inherently 
probabilistic behaviour, as well as for abstracting complexity away from 
deterministic ones. Thus, it is a long-standing interest of the verification 
community to develop logics for describing properties concerning realiability of 
software systems and to devise verification algorithms for these properties on
Markov chains and their related generalisations, such as Markov decision 
processes~\cite{mdpBellman,mdpPuterman}.

One well-known such generalisation is motivated by how the assumption of precise knowledge of a Markov chain's transition 
relation often fails to hold. Indeed, a real-world system's dynamics are rarely 
known exactly, due to incomplete information or measurement error. The need to 
model this uncertainty and to reason about robustness under perturbations in 
stochastic systems naturally gives rise to \emph{interval Markov chains (IMCs)}. In 
this model, uncertain transition probabilities are constrained to intervals, with 
two different semantic interpretations. 
Under the \emph{once-and-for-all} interpretation, the given interval Markov chain 
is seen as representing an uncountably infinite collection of Markov chains 
refining it, and the goal is to determine whether some (or alternatively, all) refinements satisfy 
a given property. In contrast, the \emph{at-every-step} interpretation exhibits 
a more game-theoretic flavour by allowing a choice over the outgoing transition 
probabilities prior to every move. The goal is then to determine strategies which 
optimise the probability of some property being satisfied. 
Originally introduced in~\cite{imcIntro}, interval Markov chains
have recently elicited considerable attention: see for example references~\cite{sen},
\cite{tah} and \cite{rasto}, which study the complexity of model checking branching-
and linear-time properties, as well as~\cite{refinement}, where the focus is
on consistency and refinement.

While IMCs are very natural for modelling uncertainty in stochastic dynamics,
they lack the expressivity necessary to capture dependencies between transition
probabilities arising out of domain-specific knowledge of the
underlying real-world system. Such a dependency could state for example that, 
although the probabilities of some set of transitions are only known to lie within a
given interval, they are all identical. Disregarding this information and studying 
only a dependence-free IMC is impractical, as allowing these transitions to vary 
independently of one another results in a vastly over-approximated space of 
possible behaviours.

Therefore, in the present paper we propose \emph{augmented interval Markov chains
(AIMCs)}, a generalisation of IMCs which allows for dependencies of this 
type to be described. We study the effect of this added expressivity through the 
prism of the (existentially quantified) reachability problem under the once-and-for-all interpretation. 
Our results are the following. First, we show that the full problem is hard
for both the famous square-root sum problem (Theorem~\ref{thm: sqrthardness}) 
and for the class $\np$ (Theorem~\ref{thm: qualNPc}). The former hardness is 
present even when the underlying graph structure is known and acyclic,
whilst the latter arises even in the qualitative subproblem when transition intervals are
allowed to include zero, rendering the structure uncertain. Second, assuming 
known structure, we show the approximate reachability problem to be in $\np$ 
(Theorem~\ref{thm: appx}). Third, we show that the restriction of the reachability 
problem to a constant number of uncertain (i.e. interval-valued) transitions is 
in~$\ptime$ (Theorem~\ref{thm: fixed}).

\section{Preliminaries}

\subsection{Markov chains}

A \emph{discrete-time Markov chain} or simply \emph{Markov chain (MC)} 
is a tuple $M = (V,\delta)$ which consists of a finite set of 
\emph{vertices} or \emph{states} $V$ and a \emph{one-step transition 
function} $\delta : V^2\rightarrow [0,1]$ such that for all $v\in V$, 
we have $\sum_{u\in V} \delta(v, u) = 1$. For the purposes of 
specifying Markov chains as inputs to decision problems, we will 
assume $\delta$ is given by a square matrix of rational numbers. The 
transition function gives rise to a probability measure on $V^\omega$ 
in the usual way. We denote the probability of reaching a vertex $t$ 
starting from a vertex $s$ in $M$ by 
$\mathbb{P}^M(s\reach t)$. The \emph{structure} of $M$ is 
its underlying directed graph, with vertex set $V$ and edge set 
$E=\{ (u,v)\in V^2 : \delta(u,v)\neq 0\}$. Two Markov chains with the 
same vertex set are said to be \emph{structurally equivalent} if their 
edge sets are identical.

An \emph{interval Markov chain (IMC)} generalises the notion of a 
Markov chain. Formally, it is a pair $(V,\Delta)$ comprising a vertex
set $V$ and a transition function $\Delta$
from $V^2$ to the set $\mathit{Int}_{[0,1]}$ of intervals contained in 
$[0,1]$. For the purposes of representing an input IMC, we will assume 
that each transition is given by a lower and an upper bound, together 
with two boolean flags indicating the strictness of the inequalities.
A Markov chain $M=(V,\delta)$ is said to \emph{refine} and interval 
Markov chain $\mathcal{M}=(V,\Delta)$ with the same vertex set if $\delta(u, v)\in\Delta(u, v)$ 
for all $u,v\in V$. We denote by $[\mathcal{M}]$ the set of Markov 
chains which refine $\mathcal{M}$. An IMC's structure is said to be 
\emph{known} if all elements of $[\mathcal{M}]$ are structurally 
equivalent. Moreover, if there exists some $\epsilon>0$ such that for
all $M=(V,\delta)\in[\mathcal{M}]$ and all $u,v\in V$, either $\delta(u,v)=0$
or $\delta(u,v)>\epsilon$, then the IMC's structure is $\epsilon$\emph{-known}. 
An IMC can have known structure but not $\epsilon$-known structure 
for example by having an edge labelled with an open interval whose lower 
bound is $0$.

An \emph{augmented interval Markov chain (AIMC)} generalises the 
notion of an IMC further by equipping it with pairs of edges whose 
transition probabilities are required to be identical. Formally, an 
AIMC is a tuple $(V, \Delta, C)$, where $(V, \Delta)$ is an IMC and 
$C\subseteq V^4$ is a set of \emph{edge equality constraints}. A 
Markov chain $(V, \delta)$ is said to refine an AIMC $(V, \Delta, C)$ 
if it refines the IMC $(V, \Delta)$ and for each $(u, v, x, y)\in C$,
we have $\delta(u, v) = \delta(x, y)$. We extend the notation 
$[\mathcal{M}]$ to AIMCs for the set of Markov chains refining 
$\mathcal{M}$.

The \emph{reachability problem} for AIMCs is the problem of deciding, 
given an AIMC $\mathcal{M}=(V, \Delta, C)$, an initial vertex 
$s\in V$, a target vertex $t\in V$, a threshold $\tau\in[0,1]$ and a 
relation $\sim \in \{\leq,\geq\}$, whether there exists 
$M\in[\mathcal{M}]$ such that $\mathbb{P}^M(\mbox{$s\reach t$})\sim
\tau$. The \emph{qualitative} subproblem is the restriction of the 
reachability problem to inputs where $\tau\in\{0, 1\}$.

Finally, in the \emph{approximate reachability problem}, we are given 
a (small) rational number $\varepsilon$ and a reachability problem 
instance. If $\sim$ is $\geq$, our 
procedure is required to accept if there exists some refining Markov 
chain with reachability probability greater than $\tau+\varepsilon/2$, it is 
required to reject if all refining Markov chains have reachability 
probability less than $\tau-\varepsilon/2$, and otherwise it is allowed to 
do anything. Similarly if $\sim$ is $\leq$.
Intuitively, this is a promise problem: in the given instance 
the optimal reachability probability is guaranteed
to be outside the interval $[\tau-\varepsilon/2,\tau+\varepsilon/2]$.

\subsection{First-order theory of the reals}

We denote by $\lang$ the first-order language $\re\langle+,\times,0,1,
<,=\rangle$. Atomic formulas in this language are of the form 
$P(x_1,\dots,x_n) = 0$ and $P(x_1,\dots,x_n) > 0$ for $P\in\zed
[x_1,\dots,x_n]$ a polynomial with integer coefficients. 
We denote by $\thr$ \emph{the first-order theory of the reals}, that
is, the set of all valid sentences in the language $\lang$. Let
$\threxst$ be the \emph{existential first-order theory of the reals},
that is, the set of all valid sentences in the existential fragment of 
$\lang$. A celebrated result \cite{Tarski51} is that $\lang$
admits quantifier elimination: each formula $\phi_1(\bar{x})$ in
$\lang$ is equivalent to some effectively computable formula
$\phi_2(\bar{x})$ which uses no quantifiers. This immediately entails
the decidability of $\thr$. 
Tarski's original result had non-elementary complexity, but 
improvements followed, culminating in the detailed analysis of 
\cite{Renegar}:
\begin{theorem}\label{thm: fo}
\begin{enumerate}
\item $\thr$ is complete for $\dblexptime$. 
\item $\threxst$ is decidable in $\pspace$. 
\item If $m\in\nat$ is a fixed constant and we 
consider only existential sentences where the number of variables 
is bounded above by $m$, then validity is decidable in $\ptime$.
\end{enumerate}
\end{theorem}
We denote by $\exists\re$ the class, introduced in~\cite{Schaefer11}, which
lies between $\np$ and $\pspace$ and comprises all problems reducible in polynomial time 
to the problem of deciding membership in $\threxst$.

\subsection{Square-root sum problem}
The \emph{square-root sum} problem is the decision problem where,
given $r_1,\dots,r_m,k\in\nat$, one must determine whether
$\sqrt{r_1} + \dots + \sqrt{r_m} \geq k$.
Originally posed in~\cite{sqrtPosed}, this problem arises naturally in computational
geometry and other contexts involving Euclidean distance. Its exact complexity
is open. 
Membership in $\pspace$ is straightforward via a reduction to the 
existential theory of the reals. Later this was sharpened in~\cite{sqrtUpperBound}
to $\posslp$, the complexity class whose complete problem is deciding whether a 
division-free arithmetic circuit represents a positive number. This class was introduced and
bounded above by the fourth level of the counting hierarchy $\ch$ in the same paper.
However, containment of the square-root sum problem in $\np$ is a long-standing 
open question, originally posed in~\cite{sqrtOpen}, and the only obstacle to
proving membership in $\np$ for the exact Euclidean travelling salesman
problem. This highlights a difference between the familiar integer model
of computation and the Blum-Shub-Smale Real RAM model~\cite{blum}, under which
the square-root sum is decidable in polynomial time~\cite{sqrtTiwari}.
See also~\cite{sqrtBackground} for more background.

\section{Qualitative case}

In this section, we will focus on the qualitative reachability problem
for AIMCs. We show that, whilst membership in $\ptime$ is straightforward 
when the underlying graph is known, uncertainty in the structure renders
the qualitative problem $\np$-complete.

\begin{theorem}\label{thm: qualP}
The qualitative reachability problem for AIMCs with
known structure is in $\ptime$.
\end{theorem}
\begin{proof}
Let the given AIMC be $\mathcal{M}$ and $s,t$ the initial and
target vertices, respectively. Since the structure $G=(V,E)$ of 
$\mathcal{M}$ is known, the qualitative reachability problem can be 
solved simply using standard graph analysis techniques on $G$. More 
precisely, for any $M\in[\mathcal{M}]$, $\mathbb{P}^M(s\reach t) = 1$
if and only if there is no path in $G$ which starts in $s$, does not enter $t$
and ends in a bottom strongly connected component which does not contain $t$. Similarly, 
$\mathbb{P}^M(s\reach t) = 0$ if and only if there is no path 
from $s$ to $t$ in $G$.
\end{proof}

\begin{theorem}\label{thm: qualNPc}
The qualitative reachability problem for AIMCs is $\np$-complete.
\end{theorem}
\begin{proof}
Membership in $\np$ is straightforward. The equivalence classes of 
$[\mathcal{M}]$ under structure equivalence are at most $2^{n^2}$, 
where $n$ is the number of vertices, since for each pair $(u,v)$ of 
vertices, either an edge $(u,v)$ is 
present in the structure or not. This upper bound is exponential in 
the size of the input. Thus, we can guess the structure of the Markov 
chain in nondeterministic polynomial time and then proceed to solve an 
instance of the qualitative reachability problem on an AIMC with known
structure in polynomial time by Theorem~\ref{thm: qualP}.

We now proceed to show $\np$-hardness using a reduction from 3-SAT.
Suppose we are given a propositional formula $\varphi$ in 3-CNF:
\begin{displaymath}
\varphi \equiv \varphi_1\wedge\varphi_2\wedge\dots\wedge\varphi_k,
\end{displaymath}
where each clause is a disjunction of three literals:
\begin{displaymath}
\varphi_i \equiv l_{i, 1} \vee l_{i, 2} \vee l_{i, 3}.
\end{displaymath}
Let the variables in $\varphi$ be $x_1,\dots,x_m$. 

Let $\mathcal{M}=(V,\Delta,C)$ be the following AIMC, also depicted in 
Figure~1. The vertex set has $3m+k+3$ vertices:
\begin{align*}
V & = \{x_1,\dots,x_m,\overline{x_1},\dots,\overline{x_m}\} \\
& \cup \{\varphi_1,\dots,\varphi_k\} \\
& \cup \{S, F\}, \\
& \cup \{v_0,\dots,v_m\}
\end{align*}
that is, one vertex for each possible literal over the given 
variables, one vertex for each clause, two special sink vertices $S,F$
(\emph{success} and \emph{failure}) and $m+1$ auxiliary vertices.
Through a slight abuse of notation, we use $x_i,\overline{x_i}$ to
refer both to the literals over the variable $x_i$ and to their 
corresponding vertices in $\mathcal{M}$, and similarly, 
$\varphi_i$ denotes both the clause in the formula and its 
corresponding vertex. 

The transitions are the following. For all $i\in\{1,\dots,m\}$, we 
have:
\begin{align*}
& \Delta(v_{i-1}, x_i) = \Delta(v_{i-1}, \overline{x_i}) =
\Delta(x_i, v_i) =\\
& \Delta(x_i, F) = \Delta(\overline{x_i}, F) =
\Delta(\overline{x_i}, v_i) = [0,1].
\end{align*}
For all $i\in\{1,\dots,k\}$ and $j\in\{1,\dots,3\}$, we have:
\begin{align*}
\Delta(\varphi_i, l_{i, j}) = [0,1].
\end{align*}
For all $i\in\{1,\dots,k\}$, 
\begin{align*}
\Delta(v_m, S) = \Delta(v_m, \varphi_i) =
\left[\frac{1}{k+1},\frac{1}{k+1}\right].
\end{align*}
Finally, $\Delta(S, S) = \Delta(F, F) = [1,1]$. For all other pairs
of vertices $u, v$, we have $\Delta(u, v) = [0,0]$.

The edge equality constraints are:
\begin{displaymath}
C = \bigcup_{i=1,\dots,m}
\{ (v_{i-1}, x_i, x_i, v_i) , 
(v_{i-1}, x_i, \overline{x_i}, F) \}
\end{displaymath}

Intuitively, the sequence of `diamonds' comprised by 
$v_0,\dots,v_m$ and the vertices corresponding to literals is a
\emph{variable setting gadget}. Choosing transition probabilities 
$\delta(v_{i-1}, x_i) = \delta(x_i, v_i) = 1$, and hence necessarily
$\delta(x_i, F) = 0$, corresponds to setting $x_i$ to true, whereas
$\delta(v_{i-1}, \overline{x_i}) = \delta(\overline{x_i}, v_i) = 1$
and $\delta(\overline{x_i}, F) = 0$ corresponds to setting $x_i$ to 
false. On the other hand, the branching from $v_m$ into 
$\varphi_1,\dots,\varphi_k$ and the edges from clauses to their 
literals makes up the \emph{assignment testing gadget}. Assigning
non-zero probability to the edge $(\varphi_i, l_{i, j})$ corresponds
to selecting the literal $l_{i, j}$ as witness that the clause 
$\varphi_i$ is satisfied.

Formally, we claim that there exists a Markov chain $M\in[\mathcal{M}]$
such that $\mathbb{P}^M(v_0\reach S)=1$ if and only if $\varphi$ is
satisfiable. 

Suppose first that $\varphi$ is satisfiable and choose some satisfying
assignment $\sigma:\{x_1,\dots,x_m\}\rightarrow\{0,1\}$. 
Let $M=(V,\delta)\in[\mathcal{M}]$ be the refining Markov chain which 
assigns the following transition probabilities to the interval-valued 
edges of $\mathcal{M}$. First, let
\begin{align*}
\delta(v_{i-1}, x_i) = \delta(x_i, v_i) = 
\delta(\overline{x_i}, F) = \sigma(x_i), \\
\delta(v_{i-1}, \overline{x_i}) = \delta(\overline{x_i}, v_i) = 
\delta(x_i, F) 
= 1-\sigma(x_i)
\end{align*}
for all $i\in\{1,\dots,m\}$.
Second, for each clause $\varphi_i$, choose some literal $l_{i,j}$ 
which is true under $\sigma$ and set $\delta(\varphi_i, l_{i,j})=1$ 
and consequently $\delta(\varphi_i, l)=0$ for the other literals $l$. 
Now we can observe that the structure of $M$ has two bottom 
strongly-connected components, namely $\{S\}$ and $\{F\}$, and 
moreover, $F$ is unreachable from $v_0$. Therefore, 
$\mathbb{P}^M(v_0\reach S)=1$.

Conversely, suppose there exists some $M=(V,\delta)\in[\mathcal{M}]$ 
such that $\mathbb{P}^M(v_0\reach S)=1$. We will prove that $\varphi$ 
has a satisfying assignment. For each $i\in\{1,\dots,m\}$, write 
\begin{align*}
p_i = \delta(v_{i-1}, x_i) = \delta(x_i, v_i) 
= \delta(\overline{x_i}, F),\\
1-p_i = \delta(v_{i-1}, \overline{x_i}) = \delta(\overline{x_i}, v_i) = 
\delta(x_i, F).
\end{align*}
Notice that 
\[
\mathbb{P}^M(v_0x_1F^\omega)=
\mathbb{P}^M(v_0\overline{x_1}F^\omega) = p_1(1-p_1),
\]
so we can conclude $p_1\in\{0,1\}$, otherwise 
$\mathbb{P}^M(v_0\reach S)\neq 1$, a contradiction. If $p_1 = 1$, 
then 
\[
\mathbb{P}^M(v_0x_1v_1x_2F^\omega) =
\mathbb{P}^M(v_0x_1v_1\overline{x_2}F^\omega) = p_2(1-p_2),
\]
whereas if $p_1 = 0$, then
\[
\mathbb{P}^M(v_0\overline{x_1}v_1x_2F^\omega) =
\mathbb{P}^M(v_0\overline{x_1}v_1\overline{x_2}F^\omega) = p_2(1-p_2).
\]
Either way, we must have $p_2\in\{0,1\}$ to ensure 
$\mathbb{P}^M(v_0\reach S)=1$. Unrolling this argument further shows 
$p_i\in\{0,1\}$ for all $i$. In particular, there is exactly one path 
from $v_0$ to $v_m$ and it has probability $1$. Let $\sigma$ be the truth 
assignment $x_i\rightarrow p_i$, we show that $\sigma$ satisfies 
$\varphi$. Indeed, if some clause $\varphi_i$ is unsatisfied under 
$\sigma$, then its three literals $l_{i,1},\dots,l_{i,3}$ are all
unsatisfied, so $\delta(l_{i,j},F)>0$ for all $j=1,\dots,3$. 
Moreover, for at least one of these three literals, say $l_{i,1}$,
we will have $\delta(\varphi_i,l_{i,1})>0$, so
the path $v_0\dots v_m\varphi_il_{i,1}F^\omega$ will have 
non-zero probability: 
\begin{align*}
\mathbb{P}^M(v_0\dots v_m\varphi_il_{i,1}F^\omega) = \frac{1}{k+1}
\delta(\varphi_i,l_{i,1})\delta(l_{i,1},F) \neq 0,
\end{align*}
which contradicts $\mathbb{P}^M(v_0\reach S)=1$.
Therefore, $\sigma$ satisfies $\varphi$, which completes the proof of
$\np$-hardness and of the Theorem.

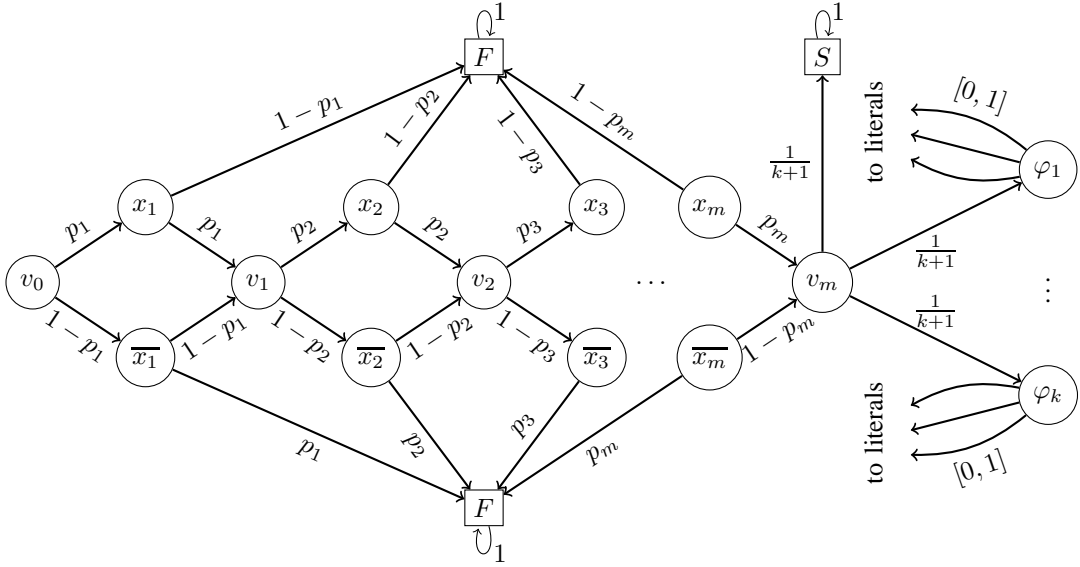
\begin{figure*}\label{fig: nphardness}
\begin{tikzpicture}[->]

\node (dx) at (1.5, 0) {};
\node (dy) at (0, 1) {};
\node (eps) at (1.5,0) {};

\node[state,minimum size=3pt] (V1) at (1.5,3) {$v_0$};
\node[state,minimum size=3pt] (V2) at ($(V1) + (dx) + (dx)$) {$v_1$};
\node[state,minimum size=3pt] (V3) at ($(V2) + (dx) + (dx)$) {$v_2$};

\node[state,minimum size=3pt] (X1) at ($(V1) + (dx) + (dy)$) {$x_1$};
\node[state,minimum size=3pt] (NX1) at ($(V1) + (dx) - (dy)$) {$\overline{x_1}$};

\node[state,minimum size=3pt] (X2) at ($(V2) + (dx) + (dy)$) {$x_2$};
\node[state,minimum size=3pt] (NX2) at ($(V2) + (dx) - (dy)$) {$\overline{x_2}$};

\node (Vi) at ($(V3) + (dx) + 0.5*(eps)$) {$\dots$};
\node[state,minimum size=3pt] (X3) at ($(V3) + (dx) + (dy)$) {$x_3$};
\node[state,minimum size=3pt] (NX3) at ($(V3) + (dx) - (dy)$) {$\overline{x_3}$};

\node[state,minimum size=3pt] (Xm) at ($(X3)+(eps)$) {$x_m$};
\node[state,minimum size=3pt] (NXm) at ($(NX3)+(eps)$) {$\overline{x_m}$};

\node[state,minimum size=3pt] (V) at ($(Vi) + (dx) + 0.5*(eps)$) {$v_{m}$};

\node[state,minimum size=3pt,shape=rectangle] (Flo) at ($(V3)-(dy)-(dy)-(dy)$) {$F$};
\node[state,minimum size=3pt,shape=rectangle] (Fhi) at ($(V3)+(dy)+(dy)+(dy)$) {$F$};

\node[state,minimum size=3pt] (Phi1) at ($(V)+2*(dx)+1.5*(dy)$) {$\varphi_1$};
\node (Phii) at ($(V)+2*(dx)$) {$\vdots$};
\node[state,minimum size=3pt] (Phik) at ($(V)+2*(dx)-1.5*(dy)$) {$\varphi_k$};

\node[state,minimum size=3pt,shape=rectangle] (S) at ($(V)+3*(dy)$) {$S$};

\node (Invis1) at ($(V) + 0.7*(dx) + 2.3*(dy)$) {};
\node[label=left:\rotatebox{90}{to literals}] (Invis2) at ($(V) + 0.7*(dx) + 2*(dy)$) {};
\node (Invis3) at ($(V) + 0.7*(dx) + 1.7*(dy)$) {};
\path[->, thick,bend right=20] (Phi1) edge node[above,sloped] {$[0,1]$} (Invis1);
\path[->, thick] (Phi1) edge node[above,sloped] {} (Invis2);
\path[->, thick,bend left=20] (Phi1) edge node[below,sloped] {} (Invis3);

\node (Invis4) at ($(V) + 0.7*(dx) - 2.3*(dy)$) {};
\node[label=left:\rotatebox{90}{to literals}] (Invis5) at ($(V) + 0.7*(dx) - 2*(dy)$) {};
\node (Invis6) at ($(V) + 0.7*(dx) - 1.7*(dy)$) {};
\path[->, thick,bend left=20] (Phik) edge node[below,sloped] {$[0,1]$} (Invis4);
\path[->, thick] (Phik) edge node[above,sloped] {} (Invis5);
\path[->, thick,bend right=20] (Phik) edge node[below,sloped] {} (Invis6);

\path[->,thick] (V1) edge node[above,sloped] {$p_1$} (X1) ;
\path[->,thick] (V1) edge node[below,sloped] {$1-p_1$} (NX1) ;
\path[->,thick] (X1) edge node[above,sloped] {$p_1$} (V2) ;
\path[->,thick] (NX1) edge node[below,sloped] {$1-p_1$} (V2) ;

\path[->,thick] (V2) edge node[above,sloped] {$p_2$} (X2) ;
\path[->,thick] (V2) edge node[below,sloped] {$1-p_2$} (NX2) ;
\path[->,thick] (X2) edge node[above,sloped] {$p_2$} (V3) ;
\path[->,thick] (NX2) edge node[below,sloped] {$1-p_2$} (V3) ;

\path[->,thick] (V3) edge node[above,sloped] {$p_3$} (X3);
\path[->,thick] (V3) edge node[below,sloped] {$1-p_3$} (NX3);
\path[->,thick] (X3) edge node[below,sloped] {$1-p_3$} (Fhi);
\path[->,thick] (NX3) edge node[above,sloped] {$p_3$} (Flo);

\path[->,thick] (Xm) edge node[above,sloped] {$p_m$} (V);
\path[->,thick] (NXm) edge node[below,sloped] {$1-p_m$} (V);
\path[->,thick] (Xm) edge node[above,sloped] {$1-p_m$} (Fhi);
\path[->,thick] (NXm) edge node[below,sloped] {$p_m$} (Flo);

\path[->,thick] (NX1) edge node[below,sloped] {$p_1$} (Flo) ;
\path[->,thick] (NX2) edge node[below,sloped] {$p_2$} (Flo) ;
\path (Flo) edge [loop below] node [right] {$1$} (Flo);

\path[->,thick] (X1) edge node[above,sloped] {$1-p_1$} (Fhi) ;
\path[->,thick] (X2) edge node[above,sloped] {$1-p_2$} (Fhi) ;
\path (Fhi) edge [loop above] node [right] {$1$} (Fhi);

\path (S) edge [loop above] node [right] {$1$} (S);
\path[->, thick] (V) edge node[left] {$\frac{1}{k+1}$} (S);
\path[->, thick] (V) edge node[below] {$\frac{1}{k+1}$} (Phi1);
\path[->, thick] (V) edge node[above] {$\frac{1}{k+1}$} (Phik);

\end{tikzpicture}
\caption{Construction used in Theorem~\ref{thm: qualNPc} for showing $\np$-hardness 
of the qualitative AIMC reachability problem. The sink $F$ is duplicated to avoid clutter.}
\end{figure*}

\end{proof}

\section{Constant number of uncertain edges}

We now shift our attention to the subproblem of AIMC reachability 
which arises when the number of interval-valued transitions is fixed,
that is, bounded above by some absolute constant. Our result is the
following.
\begin{theorem}\label{thm: fixed}
Fix a constant $N\in \nat$. The restriction of the 
reachability problem for AIMCs to inputs with at most $N$ 
interval-valued transitions lies in $\ptime$. Hence, the approximate
reachability problem under the same restriction is also in $\ptime$.
\end{theorem}
\begin{proof}
Let $\mathcal{M}=(V,\Delta,C)$ be the given AIMC and suppose we wish
to decide whether there exists $M\in[\mathcal{M}]$ such that 
$\mathbb{P}^M(s\reach t)\sim\tau$. Let $U\subseteq V$ be the set of
vertices which have at least one interval-valued outgoing transition,
together with $s$ and $t$:
\begin{displaymath}
U = \{s,t\}\cup\{ u\in V : \exists v\in V . \Delta(u,v)
\mbox{ is not a singleton} \}.
\end{displaymath}
Notice that $|U|\leq N+2=\mathit{const}$.
Write $W = V \setminus U$, so that $\{U, W\}$ is a partition of $V$.

Let $\vct{x}$ be a vector of variables, one for each interval-valued
transition of $\mathcal{M}$. For vertices $v_1,v_2$, let $\delta(v_1,
v_2)$ denote the corresponding variable in $\vct{x}$ if the 
transition $(v_1,v_2)$ is interval-valued, and the only element of
the singleton set $\Delta(v_1,v_2)$ otherwise. Let $\varphi_1$ be the
following propositional formula over the variables $\vct{x}$ which
captures the set of `sensible' assignments:
\begin{align*}
\varphi_1 \equiv & \bigwedge_{v_1\in V}\sum_{v_2\in V} \delta(v_1,v_2) = 1 \\
 \wedge & \bigwedge_{v_1,v_2\in V} \delta(v_1,v_2) \in \Delta(v_1,v_2)\cap[0,1] \\
 \wedge & \bigwedge_{(a,b,c,d)\in C} \delta(a,b) = \delta(c,d).\\
\end{align*}
There is clearly a bijection between 
$[\mathcal{M}]$ and assignments of $\vct{x}$ which satisfy $\varphi_1$.



For vertices $v_1,v_2$, use the notation $v_1\bwreach v_2$ to denote the event
\emph{`$v_2$ is reached from $v_1$ along a path consisting only of
vertices in $W$, with the possible exception of the endpoints $v_1,v_2$'}.
Notice that for all $u\in U$ and $w\in W$, $\mathbb{P}^M(w\bwreach u)$
is independent of the choice of $M\in[\mathcal{M}]$.
Denote these probabilities by $\alpha(w, u)$. They satisfy the system
\begin{displaymath}
\bigwedge_{w\in W, u\in U} \alpha(w, u) = \delta(w,u) + 
\sum_{w'\in W}\delta(w,w')\alpha(w',u),
\end{displaymath}
which is linear and therefore easy to solve with Gaussian elimination.
Thus, assume that we have computed $\alpha(w, u)\in\rat$ for all 
$w\in W$ and $u\in U$.

Similarly, for all $u_1,u_2\in U$, write $\beta(u_1,u_2)$ for the
probability of $u_1\bwreach u_2$. Notice that $\beta(u_1,u_2)$ is a 
polynomial of degree at most $1$ over the variables $\vct{x}$, given by
\begin{displaymath}
\beta(u_1,u_2) = \delta(u_1,u_2) + \sum_{w\in W} \delta(u_1, w)\alpha(w, u_2).
\end{displaymath}
Thus, assume we have computed symbolically
$\beta(u_1,u_2)\in\rat[\vct{x}]$ for all $u_1,u_2 \in U$.

Finally, for each $u\in U$, let $y(u)$ be a variable and write $\vct{y}$
for the vector of variables $y(u)$ in some order.
Consider the following formula in the existential first-order language
of the real field:
\begin{displaymath}
\varphi \equiv \exists\vct{x}\exists\vct{y}\,.\,
\varphi_1\wedge\varphi_2\wedge\varphi_3,
\end{displaymath}
where
\begin{align*}
& \varphi_2 \equiv y(t) = 1 \wedge 
\bigwedge_{u\in U\setminus\{t\}}y(u) =
\sum_{u'\in U} \beta(u, u')y(u'), \\
& \varphi_3 \equiv y(s) \sim \tau,
\end{align*}
and $\varphi_1$ is as above. Intuitively, $\varphi_1$ states that
the variables $\vct{x}$ descibe a Markov chain in $[\mathcal{M}]$,
$\varphi_2$ states that $\vct{y}$ gives the reachability 
probabilities from $U$ to $t$, and $\varphi_3$ states that the 
reachability probability from $s$ to $t$ meets the required threshold 
$\tau$. The problem instance is positive if and only if $\varphi$ is
a valid sentence in the existential theory of the reals, which is 
decidable. Moreover, the formula uses exactly
$2|U|\leq 2(N+2)=\mathit{const}$ variables, so by Theorem~\ref{thm: fo},
the problem is decidable in polynomial time, as required.
\end{proof}


Notice that removing the assumption of a constant number of 
interval-valued transitions only degrades the complexity upper bound,
but not the described reduction to the problem of checking membership
in $\threxst$. As an immediate corollary, we have:
\begin{theorem}\label{thm: ubeasy}
The reachability problem and the approximate reachability problem for 
AIMCs are in $\exists\re$.
\end{theorem}
Note that Theorem~\ref{thm: ubeasy} can be shown much more easily,
without the need to consider separately $U$-vertices and $W$-vertices
as in the proof of Theorem~\ref{thm: fixed}. It is sufficient to use
one variable per interval-valued transition to capture its transition
probability as above and one variable per vertex to express its 
reachability probability to the target. Then write down an existentially
quantified formula with the the usual system of equations for 
reachability in a Markov chain obtained by conditioning on the first 
step from each vertex. While this easily gives the $\exists\re$ upper
bound, it uses at least $|V|$ variables, so it is insufficient for 
showing membership in $\ptime$ for the restriction to a constant number 
of interval-valued transitions.

\section{Hardness for square-root sum problem}



In this section, we give a lower bound for the
AIMC reachability problem. This bound remains in place even when
the structure of the AIMC is $\epsilon$-known
and acyclic, except for the self-loops on two sink vertices.

\begin{theorem}\label{thm: sqrthardness}
The AIMC reachability problem is hard for the square-root sum problem,
even when the structure of the AIMC is $\epsilon$-known and is acyclic, 
except for the self-loops on two sink vertices.
\end{theorem}
\begin{proof}
The reduction is based on the gadget depicted in 
Figure~2. It is an AIMC with two sinks, $S$ and
$F$ (\emph{success} and \emph{failure}), each with a self-loop with 
probability $1$, and $12$ vertices: 
$\{a,b_1,\dots,b_4,c_1,\dots,c_4,d_1,d_4,e\}$.
The structure is acyclic and comprises four chains 
leading to $S$, namely, $ab_1c_1d_1eS$, $ab_2c_2S$, $ab_3c_3S$ and
$ab_4c_4d_4S$. From each vertex other than $a$ and $S$ there is also
a transition to $F$.

The probabilities are as follows. The transition $(b_3,c_3)$ has
probability $\alpha$, whilst $(b_1,c_1)$, $(b_2,c_2)$, $(b_4,c_4)$
have probability $\beta$, for rationals $\alpha,\beta$ to be specified later.
Consequently, the remaining outgoing transition to $F$ out of each $b_i$ has
probability $1-\alpha$ or $1-\beta$. The transitions $(a,b_i)$ for 
$i=1,\dots,4$ all have probability $1/4$. Finally, the transitions
$(c_1,F)$, $(c_2,F)$, $(c_3,S)$, $(c_4,F)$, $(d_1,e)$, $(d_4,S)$
and $(e,S)$ are interval-valued and must all have equal probability
in any refining Markov chain. Assign the variable $x$ to the probability
of these transitions. The interval to which these transition probabilities
are restricted (i.e. the range of $x$) is to 
be specified later. Consequently, the remaining transitions
$(c_1,d_1)$,$(d_1,F)$,$(e,F)$,$(c_2,S)$, $(c_3,F)$,$(c_4,d_4)$,$(d_4,F)$
are also interval-valued, with probability $1-x$.

Let $M$ be a positive integer large enough to ensure
\begin{align*}
x^* := \frac{3\sqrt{r}}{2M} \in (0,1).
\end{align*}
Then choose a positive integer $N$ large enough, so that
\begin{align*}
\alpha & := \frac{4M}{N} \in (0,1), \\
\beta & := \frac{16M^3}{27rN} \in (0,1), \\
p_{\mathit{opt}} & := \frac{\sqrt{r}}{N} + \frac{\beta}{4} \in (0,1). \\
\end{align*}
Now, a straightforward calculation shows
\begin{align*}
\mathbb{P}(a\reach S) & = \mathbb{P}(ab_1c_1d_1eS) + \mathbb{P}(ab_2c_2S)\\
& + \mathbb{P}(ab_3c_3S) + \mathbb{P}(ab_4c_4d_4S) \\
& = \frac{\beta x^2 (1-x)}{4} + \frac{\beta (1-x)}{4} \\
& + \frac{\alpha x}{4} + \frac{\beta x(1-x)}{4} \\
& = \frac{\alpha x - \beta x^3 + \beta}{4}.
\end{align*}
Analysing the derivative of this cubic, we see that $\mathbb{P}(a\reach S)$
increases on $[0, x^*)$, has its maximum at $x=x^*$ and then decreases 
on $(x^*,1]$. This maximum is
\begin{align*}
\frac{\alpha x^* - \beta (x^*)^3 + \beta}{4} = \frac{\sqrt{r}}{N} + \frac{\beta}{4} = p_{\mathit{opt}}.
\end{align*}
Thus, if we choose some closed interval which contains $x^*$ but not $0$ and $1$ 
to be the range of $x$, then the gadget described thus far will have 
$\epsilon$-known structure and maximum reachability probability from 
$a$ to $S$ given by $\sqrt{r}$ scaled by a constant and offset by another constant.

Now, suppose we wish to decide whether $\sqrt{r_1}+\dots+\sqrt{r_m}\geq k$
for given positive integers $r_1,\dots, r_m$ and $k$. Construct a gadget as
above for each $r_i$. The constants $\alpha, N, M$ are shared across the
gadgets, as are the sinks $S, F$, but each gadget has its own constant
$\beta_i$ in place of $\beta$, and its own copy of each non-sink vertex.
The edge equality constraints are the same as above within each gadget, and 
there are no
equality constraints across gadgets. Assign a variable $x_i$ to those 
edges in the $i$-th gadget which in the description above were labelled $x$,
and choose a range for $x_i$ as described above for $x$.
Finally, add a new initial vertex $v_0$, with $m$ equiprobable outgoing 
transitions to the $a$-vertices of the gadgets.

In this AIMC, the probability of $v_0\reach S$ is given by the
multivariate polynomial
\begin{align*}
\frac{1}{m}\sum_{i=1}^m \frac{\alpha x_i - \beta_i x_i^3 + \beta_i}{4},
\end{align*}
whose maximum value on $[0,1]^m$ is
\begin{align*}
\frac{1}{m}\sum_{i=1}^m\left(\frac{\sqrt{r_i}}{N} + \frac{\beta_i}{4}\right).
\end{align*}
Therefore, $\sqrt{r_1} + \dots + \sqrt{r_m} \geq k$ if and only if
there exists a refining Markov chain of this AIMC with
\begin{align*}
\mathbb{P}(v_0\reach S)\geq \frac{k}{mN} + \frac{1}{m}\sum_{i=1}^m\frac{\beta_i}{4},
\end{align*}
so the reduction is complete.
\end{proof}

\begin{remark}
It is easy to see that if we are given an acyclic
AIMC with the interval-valued edges labelled with variables, 
the reachability probabilities from all vertices to a single target
vertex are multivariate polynomials and can be computed symbollically
with a backwards breadth-first search from the target. Then
optimising reachability probabilities reduces to optimising the value
of a polynomial over given ranges for its variables. 

It is interesting to observe that a reduction holds in the other 
direction as well. Suppose we wish to decide whether there exist
values of $x_1\in I_1,\dots,x_n\in I_n$ such that $P(x_1,\dots,x_n)\geq\tau$
for a given multivariate polynomial $P$,
intervals $I_1,\dots,I_n\subseteq[0,1]$ and 
$\tau\in\rat$. Notice that $P$
can easily be written in the form $P(x_1,\dots,x_n) = 
\beta + N\sum_{i=1}^m \alpha_i Q_i(x_1,\dots,x_n)$, where $N>0$, $\alpha_1,\dots,
\alpha_m\in(0,1)$ are constants such that $\sum_{i=1}^m \alpha_i \leq 1$,
each $Q_i$ is a non-empty product of terms drawn from 
$\bigcup_{j=1}^n\{ x_j, (1-x_j)\}$, and $\beta$ is a (possibly negative) constant term.
For example, the monomial $-2x_1x_2x_3$ has a negative coefficient,
so rewrite it as $2(1-x_1)x_2x_3 + 2(1-x_2)x_3 + 2(1-x_3) - 2$. Do this
to all monomials with a negative coefficient, then choose an appropriately
large $N$ to obtain the desired form.

Now it is easy to construct an AIMC with two sinks $S,F$ and a designated 
initial vertex $v_0$ where the probability of $v_0\reach S$ is 
$\sum_{i=1}^m \alpha_i Q_i$. We use a chain to represent each $Q_i$, and then
branch from $v_0$ into the first vertices of the chains with distribution
given by the $\alpha_i$. There exist values of the $x_i$ in their appropriate
intervals such that $P(x_1,\dots,x_n)\geq \tau$ if and only if there exists
a refining Markov chain such that $\mathbb{P}(v_0\reach S)\geq (\tau-\beta)/N$.
\end{remark}

\begin{figure*}\label{fig: sqrthardness}
\begin{tikzpicture}[->]
\node (dx) at (4, 0) {};
\node (dy) at (0, 1.5) {};

\node[state,minimum size=3pt] (A) at (1,3) {$a$};
\node[state,minimum size=3pt] (B1) at ($(A) + 1.25*(dx) + 2*(dy)$) {$b_1$};
\node[state,minimum size=3pt] (B2) at ($(A) + (dx) + (dy)$) {$b_2$};
\node[state,minimum size=3pt] (B3) at ($(A) + (dx) - (dy)$) {$b_3$};
\node[state,minimum size=3pt] (B4) at ($(A) + 1.25*(dx) - 2*(dy)$) {$b_4$};

\node[state,minimum size=3pt] (C1) at ($(A) + 1.75*(dx) + 2*(dy)$) {$c_1$};
\node[state,minimum size=3pt] (C2) at ($(A) + 2*(dx) + (dy)$) {$c_2$};
\node[state,minimum size=3pt] (C3) at ($(A) + 2*(dx) - (dy)$) {$c_3$};
\node[state,minimum size=3pt] (C4) at ($(A) + 1.75*(dx) - 2*(dy)$) {$c_4$};

\node[state,minimum size=3pt] (D1) at ($(A) + 3*(dx) + 2*(dy)$) {$d_1$};
\node[state,minimum size=3pt] (D4) at ($(A) + 3*(dx) - 2*(dy)$) {$d_4$};

\node[state,minimum size=3pt] (E) at ($(A) + 4*(dx) + 2*(dy)$) {$e$};

\node[state,minimum size=3pt,shape=rectangle] (S) at ($(A) + 4*(dx)$) {$S$};

\node[state,minimum size=3pt,shape=rectangle] (F) at ($(A) + 1.5*(dx)$) {$F$};

\path[->, thick] (A) edge node[midway,above] {$1/4$} (B1);
\path[->, thick] (A) edge node[midway,below] {$1/4$} (B2);
\path[->, thick] (A) edge node[midway,above] {$1/4$} (B3);
\path[->, thick] (A) edge node[midway,below] {$1/4$} (B4);

\path[->, thick] (B1) edge node[midway,above] {$\beta$} (C1);
\path[->, thick] (C1) edge node[midway,above] {$1-x$} (D1);
\path[->, thick] (D1) edge node[midway,above] {$x$} (E);
\path[->, thick] (E) edge node[midway,right] {$x$} (S);

\path[->, thick] (B4) edge node[midway,below] {$\beta$} (C4);
\path[->, thick] (C4) edge node[midway,below] {$1-x$} (D4);
\path[->, thick] (D4) edge node[midway,below] {$x$} (S);

\path[->, thick] (B2) edge [bend left] node[midway,above] {$\beta$} (C2);
\path[->, thick] (C2) edge node[pos=0.4,above] {$1-x$} (S);

\path[->, thick] (B3) edge [bend right] node[midway,below] {$\alpha$} (C3);
\path[->, thick] (C3) edge node[midway,above] {$x$} (S);

\path[->, thick] (B2) edge node[pos=0.4,below,sloped] {$1-\beta$} (F);
\path[->, thick] (B3) edge node[pos=0.4,above,sloped] {$1-\alpha$} (F);
\path[->, thick] (C2) edge node[midway,below,sloped] {$x$} (F);
\path[->, thick] (C3) edge node[midway,above,sloped] {$1-x$} (F);

\path[->, thick] (B1) edge node[midway,below,sloped] {$1-\beta$} (F);
\path[->, thick] (C1) edge node[midway,right] {$x$} (F);
\path[->, thick] (B4) edge node[midway,above,sloped] {$1-\beta$} (F);
\path[->, thick] (C4) edge node[midway,right] {$x$} (F);

\path[->, thick] (D1) edge [bend left=15] node[pos=0.6,below,sloped] {$1-x$} (F);
\path[->, thick] (D4) edge [bend right=20] node[midway,above,sloped] {$1-x$} (F);
\path[->, thick] (E) edge [bend left=20] node[pos=0.6,below,sloped] {$1-x$} (F);

\path (S) edge [loop below] node [right] {$1$} (S);
\path (F) edge [loop left] node [left] {$1$} (F);

\end{tikzpicture}
\caption{Gadget for reduction from square-root sum problem to AIMC reachability.}
\end{figure*}
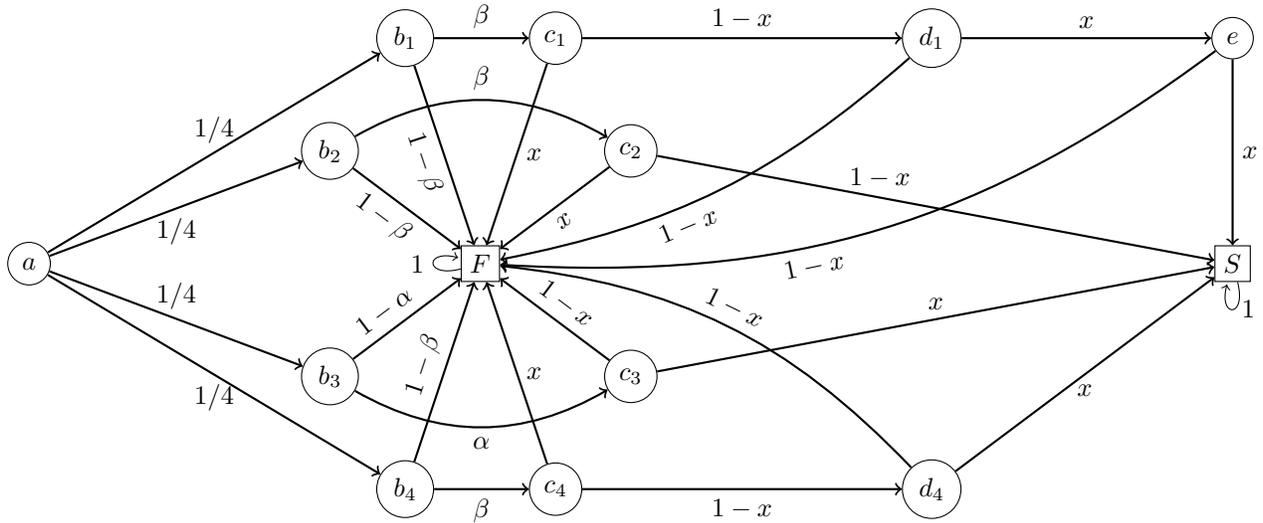

\section{Approximate case}
In this section, we focus on the approximate reachability problem for AIMCs.
To obtain our upper bound, we will use a result from~\cite{robustness}.
\begin{definition}
If $M_1=(V,\delta_1)$ and $M_2=(V,\delta_2)$ are
Markov chains with the same vertex set, then their \emph{absolute distance} is
\begin{align*}
\mathit{dist}_A(M_1,M_2) = \max_{u,v\in V} \left\{ |\delta_1(u,v)-\delta_2(u,v)| \right\}.
\end{align*}
\end{definition}

\begin{lemma}\label{lem: robust}
(Appears in~\cite{robustness}.)
Let $M_1=(V,\delta_1)$ and $M_2=(V,\delta_2)$ be structurally
equivalent Markov chains, where $n = |V|$ and for all $u,v\in V$,
we have either $\delta_1(u,v) = 0$ or $\delta_1(u, v)\geq \epsilon$.
Let also $d \leq \mathit{dist}_A(M_1,M_2)$ and fix two vertices $s,t\in V$.
Then
\begin{align*}
\left|\mathbb{P}^{M_1}(s\reach t) - \mathbb{P}^{M_2}(s\reach t)\right|
\leq
\left(1 + \frac{d}{\epsilon-d}\right)^{2n} - 1.
\end{align*}
\end{lemma}
We will also need the following well-known inequality:
\begin{lemma}\label{lem: magicIneq}
For all $x\geq -1$ and $r\in[0,1]$, we have
\begin{align*}
(1+x)^r \leq 1 + rx.
\end{align*}
\end{lemma}
Now we proceed to prove our upper bound.
\begin{theorem}\label{thm: appx}
The approximate reachability problem for AIMCs with
$\epsilon$-known structure is in $\np$.
\end{theorem}
\begin{proof}
Let $\mathcal{M}$ be the given AIMC and let $\epsilon>0$ be a lower 
bound on all non-zero transitions across all $M\in[\mathcal{M}]$.
Suppose we are solving the maximisation version of the problem: 
we are given vertices $s,t$ and a rational $\varepsilon>0$, we must 
accept if $\mathbb{P}^M(s\reach t) > \tau + \varepsilon/2$ for some
$M\in[\mathcal{M}]$ and we must reject if $\mathbb{P}^M(s\reach t) <
\tau - \varepsilon/2$ for all $M\in[\mathcal{M}]$.

Let $n$ be the number of vertices and let 
\begin{align*}
d := \epsilon\left(1 - (1+\varepsilon)^{-1/2n}\right).
\end{align*}
For each interval-valued transition, split its interval into at most
$1/d$ intervals of length at most $d$ each. For example, $[l, r]$
partitions into $[l, l+d), [l+d,l+2d),\dots, [l+kd,r]$, where $k$
is the largest natural number such that $l+kd \leq r$.
Call the endpoints defining these subintervals \emph{grid points}.
Let $\langle\mathcal{M}\rangle\subseteq[\mathcal{M}]$ be the set of Markov chains refining
$\mathcal{M}$ such that the probabilities of all interval-valued
transitions are chosen from among the grid points.
Observe that for all $M_1\in[\mathcal{M}]$, there exists 
$M_2\in\langle\mathcal{M}\rangle$ such that
$\mathit{dist}_A(M_1,M_2) \leq d$. 

Our algorithm showing membership in $\np$ will be the following.
We will choose $M\in\langle\mathcal{M}\rangle$ nondeterministically
and compute $p:=\mathbb{P}^M(s\reach t)$ using Gaussian elimination.
Then if $p\geq \tau-\varepsilon/2$, we will accept, and otherwise we
will reject. 

To complete the proof, we need to argue two points. First, that 
$\langle\mathcal{M}\rangle$ is at most exponentially large in the
size of the input, so that $M$ can indeed be guessed in nondeterministic
polynomial time. Second, that if for all $M\in\langle\mathcal{M}\rangle$
we have $\mathbb{P}^M(s\reach t) < \tau-\varepsilon/2$, then it is safe
to reject, that is, there is no $M'$ with
$\mathbb{P}^{M'}(s\reach t) \geq \tau+\varepsilon/2$. (Note that the
procedure is obviously correct when it accepts.)

To the first point, we apply Lemma~\ref{lem: magicIneq} with 
$x=-\varepsilon/(\varepsilon+1)$ and $r = 1/2n$:
\begin{align*}
(1+\varepsilon)^{-1/2n} = \left( 1 - \frac{\varepsilon}{1+\varepsilon} \right)^{1/2n}
\leq 1 - \frac{1}{2n}\frac{\varepsilon}{1+\varepsilon} 
\end{align*}
and hence,
\begin{align*}
d^{-1} = \epsilon^{-1}\frac{1}{1-(1+\varepsilon)^{-1/2n}} 
\leq \frac{1}{\epsilon}2n\frac{1+\varepsilon}{\varepsilon} 
= \frac{1}{\epsilon}2n\left(1 + \frac{1}{\varepsilon}\right).
\end{align*}
This upper bound is a polynomial in $n$, $1/\varepsilon$ and $1/\epsilon$, and hence at most exponential in the
length of the input data. Therefore, for each interval-valued transition, we can write down using only polynomially 
many bits which grid point we wish to use for the probability of that transition. Since the number of transitions
is polynomial in the length of the input, it follows that an element of $\langle\mathcal{M}\rangle$ may be specified
using only polynomially many bits, as required.

To the second point, consider $M_1, M_2\in[\mathcal{M}]$ such that
$\mathit{dist}_A(M_1, M_2)\leq d$. Then by Lemma~\ref{lem: robust},
we have
\begin{align*}
& \left|\mathbb{P}^{M_1}(s\reach t) - \mathbb{P}^{M_2}(s\reach t)\right| \\
\leq &
\left(1 + \frac{d}{\epsilon-d}\right)^{2n} - 1 \\
= & \left(\frac{\epsilon}{\epsilon(1+\varepsilon)^{-1/2n}}\right)^{2n} - 1 \\
= & \;\, \varepsilon.
\end{align*}
In other words, changing the transition probabilities by at most $d$ does not 
alter the reachability probability from $s$ to $t$ by more than $\varepsilon$.
However, recall that we chose $\langle\mathcal{M}\rangle$ in such a way that for
all $M_1\in[\mathcal{M}]$, there is some $M_2\in\langle\mathcal{M}\rangle$ with
$\mathit{dist}_A(M_1, M_2)\leq d$. In particular, if $\mathbb{P}^{M_2}(s\reach t)
< \tau - \varepsilon/2$ for all $M_2\in\langle\mathcal{M}\rangle$, then certainly
$\mathbb{P}^{M_1}(s\reach t) < \tau + \varepsilon/2$ for all $M_1\in[\mathcal{M}]$,
so it is safe to reject. This completes the proof.
\end{proof}

\bibliographystyle{alpha}
\bibliography{imc}

\end{document}